\documentclass{elsart}
\journal{New Astronomy}

\usepackage{natbib}

\usepackage{graphicx}
\usepackage{epsfig}

\usepackage{amssymb}

\newcommand{\lya}{Ly$\alpha$}
\newcommand{\zn}{$z \sim 9$ }
\newcommand{\micron}{$\mu$m}

\begin{document}

\begin{frontmatter}



\title{ZEN and the search for high--redshift galaxies}


\author{Jon Willis$^1$, Fr{\'e}d{\'e}ric Courbin$^2$, Jean-Paul Kneib$^{3}$
and Dante Minniti$^4$} 

\address{$^1$Department of Physics and Astronomy, University of
Victoria, Victoria, V8P 5C2\\ 
$^2$Laboratoire d'Astrophysique, Ecole
Polytechnique F{\'e}d{\'e}rale de Lausanne (EPFL), Observatoire, 1290
Sauverny, Switzerland \\ 
$^3$Laboratoire d'Astrophysique de
Marseille, Traverse du Siphon - B.P.8 13376 Marseille Cedex 12,
France\\ 
$^4$Departamento de Astronom{\'i}a y Astrof{\'i}sica, Pontificia
Universidad Cat{\'o}lica de Chile,  Vicu{\~n}a Mackenna
4860 Casilla 306, Santiago 22, Chile.\\
E-mail: jwillis@uvic.ca}

\begin{abstract}

We present the ZEN (z equals nine) survey: a deep, narrow $J$-band
search for proto-galactic \lya\ emission at redshifts \zn. In the
first phase of the survey, dubbed ZEN1, we combine an exceptionally
deep image of the Hubble Deep Field South, obtained using a narrow
band filter centred on the wavelength 1.187{\micron}, with existing
deep, broad band images covering optical to near infrared
wavelengths. Candidate \zn \lya-emitting galaxies display a
significant narrow band excess relative to the $J_s$--band that are
undetected at optical wavelengths. We detect no sources consistent
with this criterion to the 90\%\ point source flux limit of the NB
image, $F_{NB} = 3.28 \times 10^{-18} \rm \, ergs \, s^{-1} \,
cm^{-2}$. The survey selection function indicates that we have sampled
a volume of approximately $340 \, h^{-3} \, \rm Mpc^3$ to a \lya\
emission luminosity of $10^{43} h^{-2} \rm \, ergs \, s^{-1} $. When
compared to the predicted properties of \zn galaxies based upon no
evolution of observed $z \sim 6$ \lya-emitting galaxies, the `volume
shortfall' of the current survey, i.e. the volume required to detect
this putative population, is a factor of at least 8 to 10. We also
discuss continuing narrow $J$-band imaging surveys that will reduce
the volume shortfall factor to the point where the no-evolution
prediction from $z \sim 6$ is probed in a meaningful manner.

\end{abstract}

\end{frontmatter}

\section{Introduction}
\label{sec:intro}

Determining during which epoch and under what conditions the first
galaxies and stars formed is a key goal of galaxy evolution
studies. Observations of distant galaxies currently extend to
redshifts $z \sim 7$ (Kneib et al. 2004). In addition, a number of
studies have demonstrated that selecting faint galaxies displaying
$i-z$ colours characteristic of the Lyman Break signature at redshifts
$z \sim 6$ is an effective method to identify significant ($>10$
galaxies) samples of high-redshift galaxies (Bouwens et al. 2004a;
Dickinson et al. 2004; Stanway et al. 2004; Malhotra et al. 2005). The
addition of a narrow band (NB) filter to such broad band observations
robustly identifies \lya-emitting galaxies at redshifts that place the
emission line in the filter bandpass (Hu et al. 2004; Kodaira et
al. 2003; Rhoads et al. 2004). In the case of \lya-emitting galaxies
at $z > 5$, NB filters can exploit narrow spectral regions free from
bright night sky emission lines. The main benefit of this approach is
that the spectral location and profile of the \lya\ emission line can
be identified in a relatively straightforward manner -- leading to
samples of galaxies with very high spectral completeness (e.g. Hu et
al. 2004). The main drawback of course is that the narrow spectral
region viewed by the NB filter greatly reduces the volume sampled
compared to a broad band survey of comparable depth and field.

The properties of individual $z\sim6$ galaxies now provide relatively
detailed constraints upon the the very earliest epochs of star
formation (SF) \---\ pointing to an earlier epoch of very intense SF
just beyond the redshift limit of current, optical surveys. Two of the
brightest galaxies known at $z \sim 6$ have been detected at
3.6\micron\ and 4.5\micron\ using the {\it Spitzer} infrared space
telescope (Eyles et al. 2005). Analysis of the rest-frame UV to
optical spectral energy distribution (SED) of these galaxies indicates
the presence of stellar populations of masses $\sim 3 \times
10^{10}$~M$_\odot$ and ages $\sim 400$~Myr. Within the cosmological
model assumed in these proceedings (see below) the epoch corresponding
to the onset of SF in these galaxies is $z \sim 9$.

A key factor that determines the visibility of \lya-emitting galaxies
at $z > 6$ is the ionisation state of the inter-galactic medium (IGM).
As the neutral fraction of uniformly distributed hydrogen gas in the
Universe increases beyond 1 part in $10^5$ the associated optical
depth at rest wavelengths blueward of 1216\AA\ exceeds unity (Gunn and
Peterson 1965).  Observations of ``dark regions'' \---\ consistent
with an IGM optical depth $\tau > 1$ \---\ in redshift $z>6.2$ QSOs
suggest that we may be witnessing the onset of neutrality in the IGM
at these redshifts (Becker et al. 2001; Fan et al. 2002). As the IGM
neutral fraction increases further, the absorption profile develops a
strong damping wing extending to rest wavelengths $\lambda > 1216$\AA\
\---\ potentially absorbing the \lya\ feature in high-redshift SF
galaxies. However, the observation of a \lya-emitting galaxy at
redshift $z=6.56$ (Hu et al. 2002) is not inconsistent with
observation of a possible GP effect in lower redshift QSOs. Haiman
(2002), Santos (2004) and Barton et al. (2004) describe the ionising
effect of a star forming galaxy embedded in a neutral IGM and note
that, depending upon the exact assumptions made regarding the mass and
star formation properties of the source and the physical conditions
present in the IGM, the galaxy will form a local H{\small II} region
of sufficient size to permit transmission of a partially attenuated
\lya\ line and associated continuum.

The visibility of \lya\ emission in high-redshift SF galaxies is
important as, although $z>7$ galaxies will appear as continuum
drop-out sources, e.g. $z-J$ (Bouwens et al. 2004b), confirming the
redshift of such faint candidates ($H>27$) on the basis of continuum
features observed in deep near infrared (NIR) spectra is challenging
to the point of being impractical. Therefore, a NIR NB survey for
\lya-emitting galaxies at $z>7$ has the potential to detect
\lya-bright galaxies for which obtaining a spectroscopic redshift via
follow-up NIR spectroscopy using 8--10m class telescopes is a
realistic goal.  In the remainder of these proceedings we describe a
dedicated search for high--redshift star forming galaxies, employing
an extension of broad and narrow--band selection techniques applied at
optical wavelengths to the NIR wavelength regime. In particular we
focus upon the application of a narrow $J$--band filter centred at
$\lambda=1.187${\micron} to detect the signature of {\lya} emitting
galaxies located about a redshift $z=8.8$ (termed \zn in the following
text).

In these proceedings we adopt a cosmological model described by the
parameters $\Omega_{\rm M}=0.3$, $\Omega_\Lambda = 0.7$, $h = \rm H_0
/ 100$~kms$^{-1}$~Mpc$^{-1} = 0.7$. All magnitudes are quoted using
the AB system.

\section{Constructing the experiment: going deep and rejecting interlopers}

We assume that a redshift \zn \lya--emitting galaxy will display a
significant narrow band excess relative to the $J$--band, in addition
to displaying a continuum break consistent with almost complete
attenuation of photons at rest frame $\lambda < 1216${\AA}. In order
to generate an effective survey for such sources certain additional
factors must be considered:
\begin{enumerate}

\item{NIR continuum imaging data must achieve a limiting depth of $\rm
AB \ge 25.5$. The brightest emission line galaxies confirmed at
redshifts $z=5.7$ display AB magnitudes $z^{\prime} \approx 24.5-25$
(Hu et al. 2004). The additional distance modulus between a redshift
$z=5.7$ and $z=8.8$ results in a relative dimming term of 1
magnitude.}

\medskip

\item{Optical imaging data must reach a limiting depth typically 1.5
magnitudes fainter than NIR data. Early--type galaxies located at
redshifts $z \sim 2$ can generate a spectral discontinuity between
optical and NIR continuum bands of amplitude $D \sim 1.5$ mag.
(Stanway et al 2004b). Failure to identify the continuum break
directly could lead to the mis-identification of redshifted [OII]3727
emission in such sources as candidate \zn \lya\ emission.}

\end{enumerate}

Following these considerations the Hubble Deep Field South (HDFS;
Williams et al. 2000) Wide field Planetary Camera 2 (WFPC2) apex
pointing ($\alpha=22^h32^m55^s.64, \,
\delta=-60^\circ33^\prime05^{\prime\prime}.01$, J2000) was selected as
the target field in order to exploit the high quality of optical to
NIR image data available for the field. In particular, the combination
of HDFS WFPC2 and Very Large Telescope (VLT) Infrared Spectrometer And
Array Camera (ISAAC; Moorwood 1997) observations of the field
provide images to typical depths AB=28 and AB=26 in optical and NIR
bandpasses respectively (Labb{\'e} et al. 2003).

\begin{figure}[h]
\centerline{\psfig{file=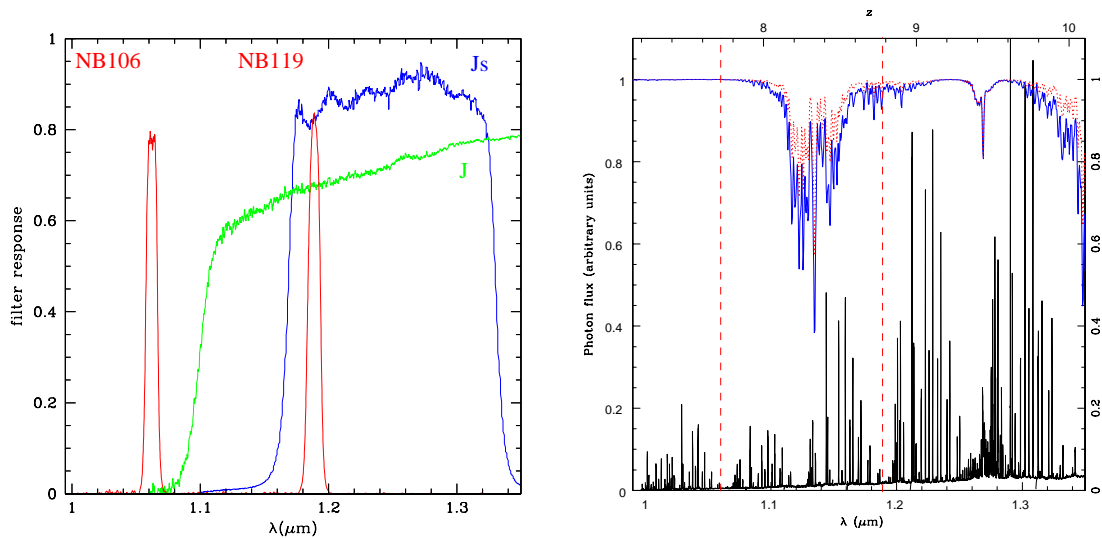,width=5.8in}}
\caption{Left panel: Filter response functions for the NB106, NB119,
$J$, and $J_s$ VLT/ISAAC filters. Right panel: The central wavelength
locations of the NB106 and NB119 filters (vertical red dotted line)
are compared to the NIR night sky emission spectrum (solid back line;
Rousselot et al. 2000) and the atmospheric transmission measured at
two epochs to illustrate temporal transmission variations (red and
blue solid lines).}
\label{fig:filters}
\end{figure}

VLT/ISAAC further provides a suitable combination of narrow and broad
band filters for the ZEN survey: the NB119 filter (see Figure
\ref{fig:filters}) is well placed within the response of both the $J$ and
$J_s$ NIR bandpasses. In addition the NB119 filter samples a spectral
region free from both strong atmospheric emission and absorption
features. The NB119 filter has an effective width of 89.5{\AA} and is
centred at a wavelength 1.187{\micron} (corresponding to the location
of {\lya} emission at a redshift $z=8.76$). While the NB106 filter
samples similarly ``dark'' sky regions, it is not well placed with
respect to the broad $J$-band filters \---\ complicating the
estimation of NB excess for individual sources.

\section{The ZEN1 data set}

Narrow $J$--band observations of the HDFS WFPC2 pointing were obtained
during ESO Period 69 (May 19th to September 17th 2002) employing the
VLT/ISAAC facility. The total NB data set consists of $420 \times
300$s spatially dithered exposures. Imaging data were a) corrected for
varying pixel response using twilight sky exposures, b) sky-subtracted
having masked array regions containing objects detected above a
specified ADU level, c) corrected for both high-- and low--frequency
spatial artefacts, d) shifted to a common pixel scale and coadded
using a suitable pixel weighting and rejection algorithm. The applied
data reduction techniques are broadly similar to those described in
Labb{\'e} et al. (2003) for deep $J_sHK_s$ observations of the HDFS. A
more complete description of the NB data reduction and analysis can be
found in Willis and Courbin (2005).
\begin{figure}[h]
\centerline{\psfig{file=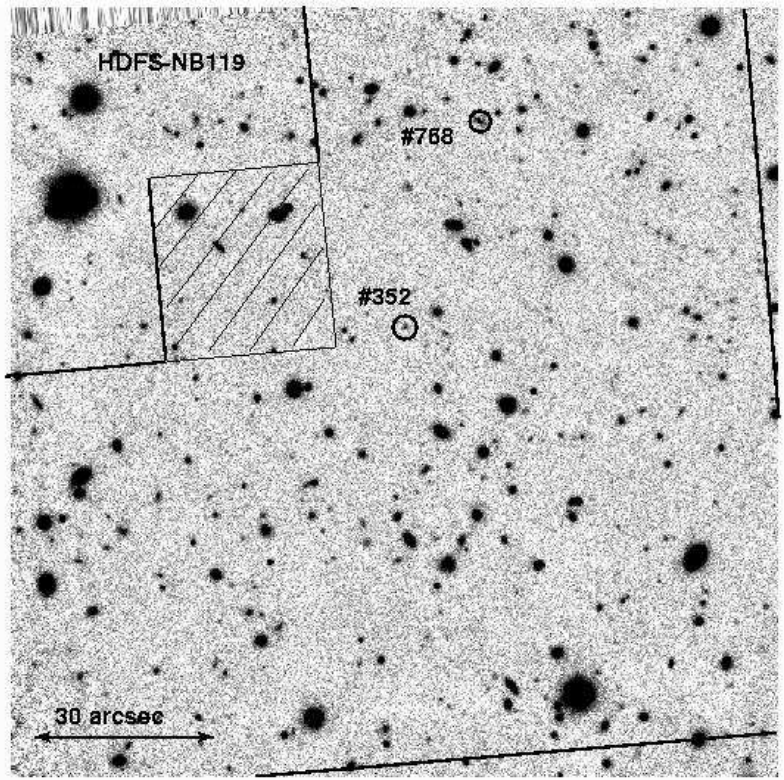,width=5.8in}}
\caption{Greyscale NB119 image of the HDFS. The image orientation is
North up and East left. The field geometry of the NIR broad band data
set is identical to the NB image. The field geometry of the HST WFPC2
field is indicated. Note that the Planetary Camera (shaded region) data
does not contribute to the final catalogue. The total field area
contributing to the NB excess catalogue is 4 arcmin$^2$. The location
of the two NB excess sources identified from the catalogue are
indicated (see text for details).}
\label{fig:image}
\end{figure}

The photometric zero point for the NB119 image was computed employing
$5^{\prime\prime}$ diameter aperture photometry of 47 bright, isolated
sources common to the NB119 and FIRES $J_s$--band images. Assuming
that the SEDs of these calibration sources display no strong
discontinuities, the $J_s$--band flux density provides an accurate
estimate of the flux density within the NB119 filter, i.e. $J_s-NB =
0$. Source detection and photometry was performed on the NB119 image
and the $0.^{\prime\prime}7$ diameter detection apertures were
transformed to the FIRES astrometric system to compute corresponding
$U_{300}B_{450}V_{606}I_{814}J_sHK_s$ magnitudes.

The flux completeness limit of the NB119 and $J_s$ images was
estimated by introducing and recovering artificial unresolved sources
within each field. Adopting the 90\%\ point source recovery threshold
as the limiting magnitude in each band generates magnitude limits of
$NB \le 25.2$ and $J_s \le 26.2$ and corresponds to integrated
signal--to--noise ratios (SNR) of 13.8 and 10 respectively.  The
magnitude limit $NB = 25.2$ corresponds to a total flux integrated
across the NB filter of $F_{NB} = 3.28 \times 10^{-18} \, \rm ergs \,
s^{-1} \, cm^{-2}$. In common with Labb{\'e} et al. (2003) we present
$0.^{\prime\prime}7$ aperture photometry throughout this paper. The
correction required to convert photometric measures computed in
$0.^{\prime\prime}7$ apertures to $5^{\prime\prime}$ apertures (which
we assume to be `total' measures) was determined to be 0.7 magnitudes
via analysis of bright stars in the NB image.

\begin{figure}[h]
\centerline{\psfig{file=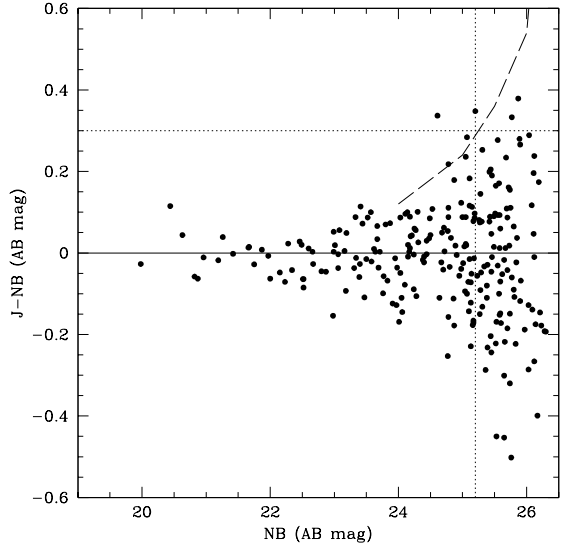,width=3.0in}}
\caption{Distribution of NB excess $J_s-NB$ versus $NB$ magnitude for
all objects in the NIR field area of the HDFS (solid points). The
solid horizontal line indicates $J_s-NB=0$. The vertical dotted line
indicates the selection criterion $NB\le25.2$. The horizontal dotted
line indicates the selection criterion $J_s-NB\ge0.3$. The dashed
curve indicates the predicted $J_s-NB$ uncertainty ($3\sigma$) as a
function of $NB$ magnitude returned by the completeness analysis.}
\label{fig:cmd}
\end{figure}

Figure \ref{fig:cmd} displays NIR narrow band excess, expressed as
$J_s-NB$, versus $NB$ magnitude for all sources extracted from the
total area covered by the optical and NIR data for the HDFS
field. Note that all NB detections have a corresponding $J_s$
detection. Sources displaying $J_s-NB \ge 0.3$ and $NB \le 25.2$ are
flagged as potential \zn emitting galaxies.

\begin{figure}[h]
\centerline{\psfig{file=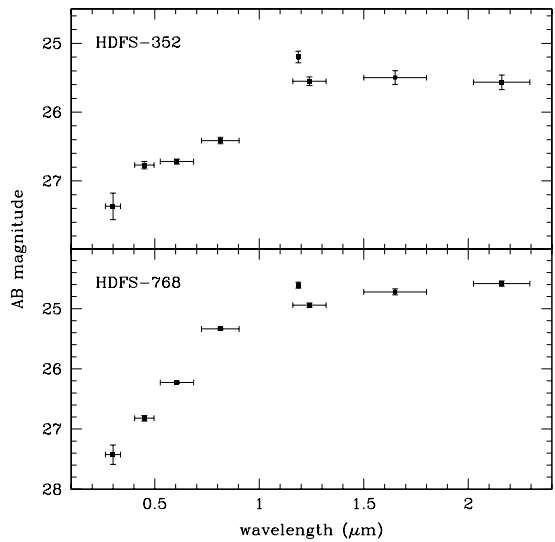,width=3.0in}}
\caption{Low resolution spectral energy distributions formed by the
multi-band photometry of the two NB excess objects identified within
the HDFS. The clear detection of each source within optical passbands
precludes each source as a candidate \zn source.}
\label{fig:photcand}
\end{figure}

Two sources are identified by the above selection criteria, HDFS--352
and HDFS--768 \---\ where the identification numbers refer to the HDFS
photometric catalogue of Labb{\'e} et al. (2003). The photometric
spectrum formed by the $U_{300}B_{450}V_{606}I_{814}J_sHK_s$ plus NB
photometry of each object is displayed in Figure
\ref{fig:photcand}. In each case the addition of deep optical
photometry is sufficient to exclude each NB excess detection as a
potential \zn source: each NB excess object is detected in all optical
bands and the photometric redshift of each source (Rudnick et
al. 2001) -- $z_{phot} = 1.54 \pm 0.06$ (HDFS--352) and $z_{phot} =
0.76_{-0.20}^{+0.04}$ (HDFS--768) -- is consistent with a NB excess
arising from redshifted H$\beta \, 4861$ and H$\alpha \, 6563$
emission respectively.
Although no candidate \zn sources are detected within the HDFS survey
area, the confirmation of two sources whose photometric redshifts are
consistent with the observed NB excess arising from narrow line
emission lends support to the assertion that the adopted selection
criteria identify faint, narrow emission line galaxies.

\section{A ZEN paradox: what is the significance of detecting no \lya\ emitters at $z=9$?}

\begin{figure}[h]
\centerline{\psfig{file=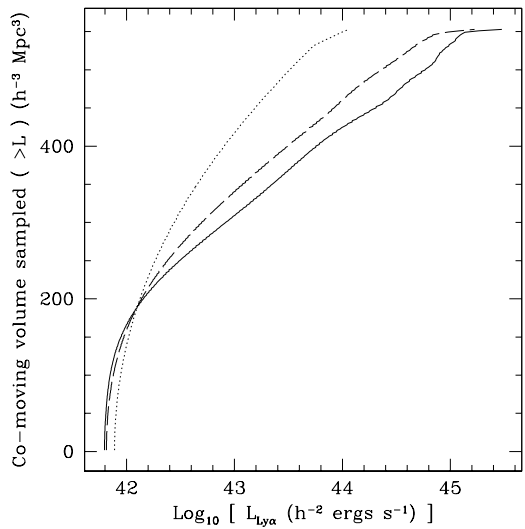,width=3.0in}}
\caption{Co-moving volume sampled as a function of \lya\ emission
luminosity. Three values of the rest frame \lya\ velocity width are
displayed; $\sigma_v = 20$ (solid line), 50 (dashed line) and 100
(dotted line) kms$^{-1}$.}
\label{fig:volsamp}
\end{figure}

The \lya\ emission line selection function generated by the NB excess
search technique is displayed in Figure \ref{fig:volsamp} in terms of
the co--moving volume sampled as a function of \lya\ emission
luminosity for three values of the rest frame velocity width of the
\lya-emitting source. For the case where the rest frame velocity width
of putative \zn sources is $\sigma_v = 50 \rm \, kms^{-1}$, the NB
excess survey area samples a co--moving volume of $340 \, h^{-3} \rm
\, Mpc^3$ to a \lya\ emission luminosity of ${\rm L_{Ly\alpha}} \ge
10^{43} \, h^{-2} \rm \, ergs \, s^{-1}$.

The uncertain physical state of the IGM at \zn complicates a detailed
comparison of model \lya-emitting populations to the region of
luminosity and volume parameter space sampled by the ZEN1 study
(e.g. see Santos 2004). A more empirical approach is to consider known
\lya-emitting sources and to transpose their properties to redshift
\zn. Hu et al. (2004) (hereafter H04) present a search for
\lya-emitting galaxies at $z \sim 5.7$ employing a narrow band filter
centred on the wavelength 8150\AA. Though the H04 study does not
present the highest redshift \lya\ sources currently known
(c.f. Kodaira et al. 2003), their study is based upon near complete
spectroscopic follow-up observations \---\ 19 out of 23 observed
sources (from a total sample of 26 candidates) were confirmed as \lya\
at $z=5.7$ \---\ and the overall design of their study is very close
to the ZEN1 survey.  H04 report a surface density of \lya\ emitting
galaxies of 0.03 arcmin$^{-2}$ to a \lya\ flux limit of $2 \times
10^{-17}$ ergs s$^{-1}$ cm$^{-2}$ from a total areal coverage of 700
arcmin$^2$.

Within our adopted cosmological model, an unresolved source at a
redshift $z=8.8$ is 2.7 times fainter than an identical source viewed
$z=5.7$. A further factor of 0.5 must be applied to account for the
different fraction of total flux measured by the $3^{\prime\prime}$
diameter apertures applied by H04 and the $0.^{\prime\prime}7$
diameter apertures employed in the current study. The H04 flux limit
therefore corresponds to an approximate limit of $3.5 \times 10^{-18}$
ergs s$^{-1}$ cm$^{-2}$ when transposed to a redshift $z = 8.8$ and
employing $0.^{\prime\prime}7$ diameter apertures, i.e., assuming that
no additional evolution occurs, the $z=5.7$ sources observed by H04
are sufficiently bright to be observed within ZEN1. Applying the
surface density of confirmed $z=5.7$ \lya\ emitting galaxies to the
areal coverage of the current NIR survey, indicates a probability to
detect a \zn galaxy of 0.12 \---\ assuming no evolution between the
two redshifts. This `volume shortfall' indicates that the current NIR
selected survey will have to be extended by more than eight times the
currently sampled area in order to realistically probe the no
evolution scenario.

The present ZEN1 survey has successfully demonstrated that the
sensitivity required to detect putative \zn \lya\ emitting galaxies
emission has been achieved for the case where no evolution occurs
between redshifts $z=5.7$ and $z=8.8$.  However, it is clear that, in
order to place stronger limits on the space density of such sources,
the areal coverage of the current study must be extended.

\section{Continuing efforts}

A number of observational programs are underway with the aim of
detecting $z>7$ \lya-emitting galaxies using the narrow $J$-band
technique described above (see contributions from Smith et al. and
Bland-Hawthorn et al. in these proceedings). We are currently
undertaking ZEN2, a NB119 survey of three massive lensing clusters
\---\ AC114, A1869 and A1835. The gravitational lens effect provides a
typical magnification across each ISAAC field of three and our
observations have been designed to reach the same unlensed luminosity
and volume sensitivity as achieved with ZEN1. Of particular interest
however, are the regions of each cluster field that correspond to
critical lines (regions of very high magnification) for sources at
\zn. These limited areas in each ISAAC image provide exceptionally
deep sightlines and offer an important `keyhole' through which to
detect exceptionally faint emitters.

Therefore, with the completion of ZEN2 and existing NIR NB surveys
(most notably DAZLE; Bland-Hawthorn et al. in these proceedings) the
pursuit of $z>7$ \lya-emitting galaxies will have reached the stage
where simple and robust galaxy evolution models can be tested
directly.


\begin{thebibliography}{99}
\frenchspacing

\bibitem{barton04}
Barton, E.~J., Dav{\'e}, R., Smith, J-D., et al., 2004, ApJL, 604, 1

\bibitem{becker01}
Becker, R.~H., Fan, X., White, R.~L., et al., 2001, AJ, 122, 2850

\bibitem{boo04a}
Bouwens, R.~J., Illingworth, G.~D., Thompson, R.~I., et al., 2004a, ApJ, 606, 25

\bibitem{boo04b}
Bouwens, R.~J., Thompson, R.~I., Illingworth, G.~D., et al., 2004b, ApJL, 616, 79

\bibitem{dick04}
Dickinson, M., Stern, D., Giavalisco, M., et al., 2004, ApJL, 600, 99

\bibitem{eyles05}
Eyles, L., Bunker, A., Stanway, E., et al., 2005, MNRAS in press [astro-ph/0302213].

\bibitem{fan02}
Fan, X., Narayanan, V.~K., Strauss, M.~A., et al., 2002, AJ, 123, 1247

\bibitem{gunn65}
Gunn, J.~E., Peterson, B.~A., 1965, ApJ, 142, 1633

\bibitem{haiman02}
Haiman, z., 2002, ApJL, 576, 1

\bibitem{hu02}
Hu, E.~M., Cowie, L.~L., McMahon, R.~G., et al., 2002, ApJL, 568, 75

\bibitem{hu04} Hu, E.~M., Cowie, L.~L., Capak, P., et al., 2004, AJ, 127, 563

\bibitem{kneib04}	
Kneib, J-P., Ellis, R.~S., Santos, M.~R., Richard, J., 2004, ApJ, 607, 697

\bibitem{koda03}
Kodaira, K. et al. 2003, PASJL, 55 17

\bibitem{labbe03}
Labb{\'e}, I., Franx, M., Rudnick, G., et al., 2003, AJ, 125, 1107

\bibitem{mal05}
Malhotra, S., Rhoads, J. E., Pirzkal, N., et al., 2005, ApJ, 626, 666

\bibitem{moorwood97}
Moorwood, A., 1997, Proc. SPIE, 2871, 1146

\bibitem{rhoads04}
Rhoads, J.~E., Xu, C., Dawson, S., et al., 2004, ApJ, 611, 59

\bibitem{rousselot00} 
Rousselot, P., Lidman, C., Cuby, J.-G., et al., 2000, A\&A, 354, 1134

\bibitem{rudnick01}
Rudnick, G., Franx, M., Rix, H-W., et al., 2001, AJ, 122, 2205

\bibitem{santos04}
Santos, M.~R., 2004, MNRAS, 349, 1137

\bibitem{stan04}
Stanway, E.~R., Bunker, A.~J., McMahon, R.~G., et al., 2004, ApJ, 607, 704

\bibitem{williams00}
Williams, R.~E., Baum, S., Bergeron, L.~E., et al., 2000, AJ, 120, 2735

\bibitem{willis05}
Willis, J.~P. and Courbin, F., 2005, MNRAS, 357, 1348

\end{thebibliography}
\end{document}